\documentclass[prl,a4paper,showpacs,twocolumn,superscriptaddress,longbibliography]{revtex4-1}
\usepackage[english]{babel}
\usepackage[T2A]{fontenc}
\usepackage{amssymb}
\usepackage{amsmath}
\usepackage{amsfonts}
\usepackage{graphicx}
\usepackage{bm}
\usepackage{natbib}
\usepackage[hypertexnames=false]{hyperref}
\usepackage{graphicx}
\usepackage{bm}
\usepackage[usenames,dvipsnames]{color}

\def\be{\begin{equation}}
\def\ee{\end{equation}}

\def\br{{\bf r}}

\def\lambdaBCS{\lambda_{\text{BCS}}}
\def\lambdaph{\lambda_{\text{ph}}}

\def\qxy{q_{\parallel}}

\renewcommand{\Im}{\mathop{\rm Im}}

\newcommand\tit[1]{\relax}

\begin{document}
\title{Superconductivity suppression in disordered films:\\
Interplay of two-dimensional diffusion and three-dimensional ballistics}

\author{Daniil S. Antonenko}
\affiliation{Skolkovo Institute of Science and Technology, Moscow 121205, Russia}
\affiliation{L. D. Landau Institute for Theoretical Physics, Chernogolovka 142432, Russia}
\affiliation{Moscow Institute of Physics and Technology, Moscow 141700, Russia}

\author{Mikhail A. Skvortsov}
\affiliation{Skolkovo Institute of Science and Technology, Moscow 121205, Russia}
\affiliation{L. D. Landau Institute for Theoretical Physics, Chernogolovka 142432, Russia}

\date{September 11, 2020}

\begin{abstract}
Suppression of the critical temperature in homogeneously disordered superconducting films is a consequence of the disorder-induced enhancement of Coulomb repulsion. 
We demonstrate that for the majority of thin films studied now this effect cannot be completely explained in the assumption of two-dimensional diffusive nature of electrons motion. 
The main contribution to the $T_c$ suppression arises from the correction to the electron-electron interaction constant coming from small scales of the order of the Fermi wavelength that leads to the critical temperature shift $\delta T_c/T_{c0} \sim - 1/k_Fl$, where $k_F$ is the Fermi momentum and $l$ is the mean free path.
Thus almost for all superconducting films that follow the fermionic scenario of $T_c$ suppression with decreasing the film thickness, this effect is caused by the proximity to the three-dimensional Anderson localization threshold and is controlled by the parameter $k_F l$ rather than the sheet resistance of the film.

\end{abstract}

\maketitle

\textbf{1. Introduction.}
The principal characteristics of a superconductor is the value of its transition temperature, $T_c$. It is usually assumed that $T_c$ is a material property and does not depend on the sample size.
However there is a strong experimental evidence of the systematic decrease of the critical temperature in disordered superconducting films with decreasing its thickness, $d$ 
(V \cite{Teplov_V_1976}, 
NbN \cite{Wang_NbN_1996, SemenovGippius_NbN_2009, NoatRodichev_NbN_2013, Makise_NbN_2015, Kang_NbN_2011, EzakiMakise_Nb_2012, ChandRaychaudhuri_NbN_2012, Brun_NbN_2020},
TiN \cite{SacepeBaturina_TiN_2008},
MoGe \cite{GraybealBeasley_MoGe_1984, Lotnyk_2017},
MoSi \cite{Fogel_MoSi_1996,Banerjee_2017},
MoC \cite{SzaboSamuely_MoC_2016}, 
WRe \cite{Raffy_WRe_1983},
InO \cite{InO-Shahar}
etc. \cite{Strongin_PbBiSnAl_1970}).
The suppression of $T_c$ becomes pronounced typically at $d\sim 10$ nm, and for the thinnest films $T_c$ may eventually vanish, marking the point of a quantum superconductor-metal or superconductor-insulator transition
\cite{Haviland_Bi_1989, Fisher1990, GantmakherDolgopolov_SIT_2010, Burmi-SIT, KKS, Sacepe2020rev}.

Depending on the underlying structure of a material, two scenarios of $T_c$ suppression --- fermionic and bosonic --- have been identified. The bosonic scenario applies to granular and/or strongly inhomogeneous superconductors with localized preformed Cooper pairs (polycrystalline TiN, amorphous InO) \cite{QSMT2001, Feigel2007, FEIGELMAN20101390, InO-preloc}, where $T_c$ signals proliferation of superconducting coherence from micro- to macro-scales.
In the fermionic scenario, relevant for structureless homogeneously disordered superconductors (NbN, MoGe, etc.), suppression of superconductivity is a consequence of the disorder-induced enhancement of electron repulsion \cite{AltshulerAronov_ContributionDisorderedMetals_1979, AltshulerAronov_book_1985}, which leads to the decrease in the effective Cooper pairing constant.  
Despite the common physical mechanism of disorder-induced $T_c$ suppression in the fermionic scenario, its description for three- and two-dimensional systems is rather different.

\emph{In the three-dimensional (3D) geometry}, enhancement of repulsion due to scattering off the impurity potential is provided by small distances, not exceeding the mean free path $l$. As a result, the whole effect can be completely described by the change in the Cooper pairing constant. The fermionic mechanism for strongly disordered 3D superconductors in the vicinity of the Anderson localization threshold ($k_F l\sim1$, where $k_F$ is the Fermi momentum) was studied by Anderson, Muttalib and Ramakrishnan \cite{AndersonMuttalibRamarkrishnan}.
They also estimated the correction to the bare electron-electron interaction constant $\lambda$ in the case of weak disorder ($k_F l\gg1$): $\delta \lambda / \lambda \sim 1 / (k_F l)^2$.
Similar expressions were reported in Refs.\  \cite{Fukuyama_Bulk_SC_suppression, Rabatin_2018}. 
This estimate can be easily obtained by cutting the 3D diffusive contribution at the ultraviolet cutoff $r\sim l$. 
However, as shown by Belitz and Kirkpatrick in their study of weak-localization correction to the conductivity  \cite{BelitzKirkpatrick_UltrasonicAttenuationDisordered_1986}, diffusive contributions in the 3D geometry are extended to the ballistic region up to the distances of the order of wavelength and have a relative order of $1/(k_F l)$ rather than $1/(k_F l)^2$. 
Similar extension of the interaction-induced contribution from the diffusive to the ballistic region is also known for the tunneling density of states, both in 2D \cite{Glazman} and 3D geometries \cite{Koulakov_QuasiballisticDOS_2000, AntSkv2020}.

Disorder-induced renormalization of the electron-phonon interaction and its impact on superconductivity were studied by Keck and Schmid \cite{KeckSchmid_SuperconductvityElectronPhononImpureMetals_1976}. 
They showed that the displacement of impurities by the lattice vibrations leads to the suppression of the interaction with longitudinal phonons and the emergence of the interaction with transverse phonons. 
An attempt to account for the impurity corrections both to the Coulomb and  electron-phonon interactions and their influence on $T_c$ was taken by Belitz with the help of the exact-eigenstates technique \cite{Belitz-correlation-gap} and by solving full Gor'kov equations in the strong-coupling regime \cite{Belitz_StrongCoupling1_1987, Belitz_StrongCoupling2_1987, Belitz_Tc_1987}.
A part of his results can be interpreted as a correction to the bare electron-electron coupling constant $\delta \lambda / \lambda \sim 1 / k_F l$. 
However, Belitz's results were called into question by Finkel'stein  \cite{Finkelstein_review_SCSuppressionDisorderedFilms_1994} by demonstrating that elastic diagrams, intimately related to the correction to the tunneling density of states \cite{MaekawaFukuyama_Localization2DSC_1982, Maekawa_UpperCriticalField2DSC_1983} and claimed to be essential by Belitz, actually do not contribute to the leading order of $T_c$ shift. 

\begin{figure}
\includegraphics[width=0.9\linewidth]{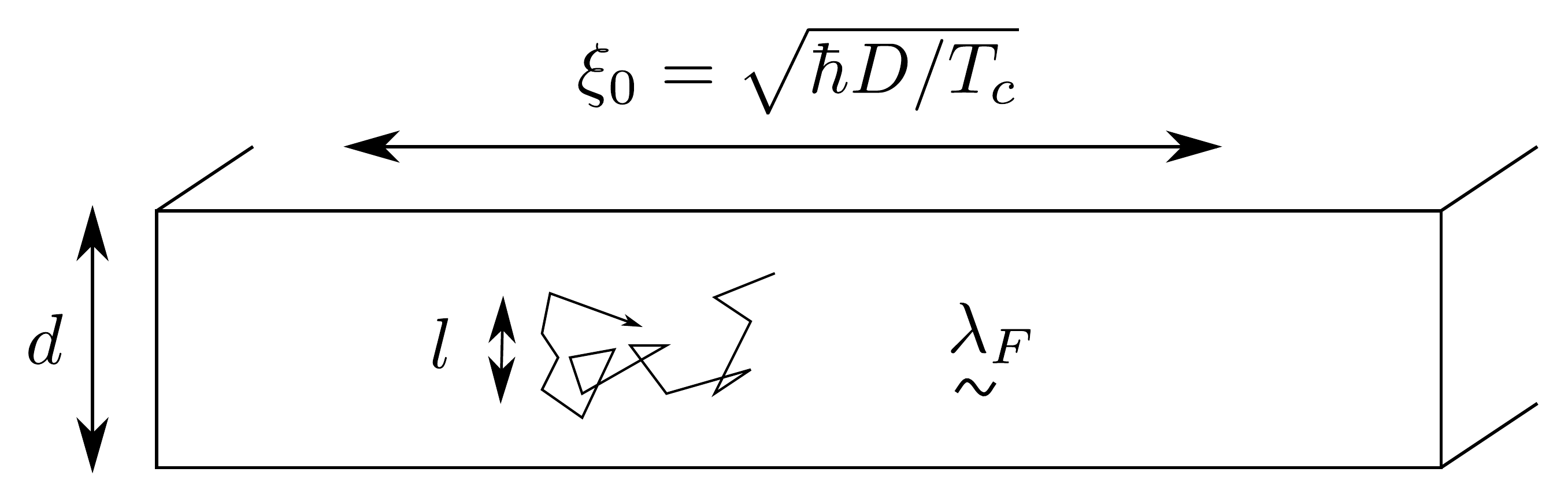}
\protect\caption{Experimentally relevant hierarchy of length scales in disordered superconducting films.} 
\label{fig:Scales}
\end{figure}

The main difference of \emph{the two-dimensional (2D) geometry} compared to the 3D case is that the renormalization effect does not boil down to the energy-independent shift of the coupling constant $\lambda$ and requires a summation of the leading logarithms. 
Conventional description of $T_c$ suppression in thin superconducting films substantially relies on \emph{2D diffusive} nature of electron motion, which is motivated by the experimentally relevant hierarchy of length scales $\lambda_F \ll l \ll d \ll \xi_0$, see Fig.~\ref{fig:Scales}.
(Here $\lambda_F$ is Fermi wavelength, $\xi_0=\sqrt{\hbar D/T_c}$ is the superconducting coherence length in the dirty limit, and $D$ is the diffusion constant.)
In this paradigm, enhancement of disorder with the decrease of the film thickness $d$ is related to the increase of the sheet resistance of the film, $R_\Box$.

The effect of $T_c$ shift due to the interplay of disorder and interaction was studied on a perturbative level in Refs.\ \cite{Ovchinnikov_FluctuationShiftOfTcSCFilms_1973, MaekawaFukuyama_Localization2DSC_1982, TakagiKuroda_AndersonLocSCTransition2D_1982, Maekawa_UpperCriticalField2DSC_1983, EbisawaFukuyamaMaekawa_SCTcDirtyFilms_1985},
where the 2D diffusive contribution to the $T_c$ shift was calculated:
\be
\label{Ovchinnikov_and_Japanese_result}
\frac{\delta T_c}{T_{c0}}
=
- \frac{\lambda}{ 3\pi g} \log^3 \frac{\hbar}{T_{c0}\tau_*}
,
\ee
where $T_{c0}$ is the critical temperature of a bulk super\-conductor, $g=h/e^2R_\Box=(2/3\pi)(k_Fl)(k_Fd)\gg1$ is the dimensionless film conductance, and $\lambda$ is the dimensionless coupling constant of the electron-electron interaction (for the screened Coulomb interaction, $\lambda = 1/2$).
The parameter $\tau_*$ is the time when diffusion becomes two-dimensional: $\tau_*=\max\{\tau,\tau_d\}$, where $\tau$ is the elastic scattering time and $\tau_d=d^2/4D$ is the time of diffusion across the film thickness \cite{Ovchinnikov_FluctuationShiftOfTcSCFilms_1973, Finkelstein_review_SCSuppressionDisorderedFilms_1994}.
In real space, the logarithm in Eq.~\eqref{Ovchinnikov_and_Japanese_result} is accumulated from the 2D diffusion from the length scale $\max(l,d)$ to the coherence length  $\xi_0$.
The correction \eqref{Ovchinnikov_and_Japanese_result}, inversely proportional to the film conductance, is conceptually similar to the weak-localization \cite{band4,GLK} and interaction-related \cite{AltshulerAronov_book_1985} corrections to the 2D conductivity, while two out of three powers of the logarithm are due to the exponential sensitivity of $T_c$ to the coupling constant~$\lambda_\text{BCS}$.

The first-order perturbative result \eqref{Ovchinnikov_and_Japanese_result} has later been generalised to the case of arbitrarily strong $T_c$ suppression by Finkel'stein, who managed to sum the leading logarithms with the help of the renormalization-group technique \cite{Finkelstein_TcSCfilms_JETP_1987, Finkelstein_review_SCSuppressionDisorderedFilms_1994}. 
The same result can be obtained by solving the self-consistency equation with an energy-dependent Cooper coupling $\lambda_{E,E'}=\lambdaBCS-\gamma_g^2\log[1/\max(E,E')\tau_*]$ \cite{FS2012}.
For the screened Coulomb interaction ($\lambda=1/2$), the nonperturbative expression for the critical temperature as a function of the dimensionless film conductance valid until superconductivity is fully suppressed is given by:
\be
\label{Fin}
\log\frac{T_c}{T_{c0}}
= 
\frac{1}{\gamma} 
-
\frac{1}{2\gamma_g} 
\log\frac{\gamma+\gamma_g}{\gamma-\gamma_g} ,
\ee
where $\gamma_g=1/\sqrt{2\pi g}$ and $\gamma=1/\log(\hbar/T_{c0}\tau_*)$.
Expression \eqref{Fin}, \emph{where $\gamma$ is considered as a fitting parameter}, was used by Finkel'stein \cite{Finkelstein_review_SCSuppressionDisorderedFilms_1994} to explain the observed dependence of $T_c$ in MoGe films  \cite{GraybealBeasley_MoGe_1984} on the film thickness, the latter being directly related to the dimensionless conductance $g$.
Since then, such an explanation of experimental data on superconductivity suppression in disordered films has become generally accepted \cite{KimRogachev_MagneticGdImpuritiesSCMoGe_2012, SzaboSamuely_MoC_2016, Banerjee_2017}.

According to Eqs.\ \eqref{Ovchinnikov_and_Japanese_result} and \eqref{Fin}, $T_c$ suppression in thin ($d\ll\xi_0$) superconducting films is entirely determined by the dimensionless sheet conductance $g$.
Such a statement perfectly fits the general theoretical framework of scaling \cite{band4}, justified by the renormalization-group analysis of the nonlinear sigma model in the 2D space \cite{Efetov-book,Fin,Burmistrov_2019}.

However, interpretation of experimental data on $T_c(d)$ dependence with the help of Eq.\ \eqref{Fin} encounters a number of significant difficulties. 
The first one is the internal inconsistency of the approach that treats $\gamma$ as a free fitting parameter.
As follows from Table \ref{tbl:films}, which contains experimental data on different films, the values of $\gamma^{-1}_\text{fit}$ obtained by fitting $T_c(d)$ dependence with the help of Eq.\ \eqref{Fin} typically lie in the interval $7\div9$.
The issue is that these values significantly exceed the theoretical estimate $\gamma^{-1}={\cal L}_d=\ln(\hbar/T\tau_d)$ (last column in Table \ref{tbl:films}), and in half of the cases exceed even the quantity ${\cal L}=\ln(\hbar/T\tau)$ (last but one column in Table \ref{tbl:films}). 
Taking into account that the perturbative shift of $T_c$, according to Eq.~\eqref{Ovchinnikov_and_Japanese_result}, is proportional to the third power of this logarithm, one can conclude that the discrepancy between the microscopic theory and the result of the fit with Eq.\ \eqref{Fin} appears to be very large. 
One can try to save the situation by pointing to the fact that $\gamma^{-1}_\text{fit}$ should also contain the contribution of 3D diffusion, but that makes the usage of Eqs.~\eqref{Ovchinnikov_and_Japanese_result} and \eqref{Fin} dubious as they were obtained under the assumption of 2D diffusion.

\begin{table}
\caption{Parameters of superconducting films \cite{com-table}:
bulk critical temperature $T_{c0}$, thickness $d$, mean free path $l$, the value of $\gamma$ obtained from fitting $T_c(g)$ dependence with Eq.~\eqref{Fin} and the values of the two logarithms:
 ${\cal L}=\log (\hbar/T_{c0} \tau)$ and ${\cal L}_d=\log (\hbar/T_{c0} \tau_d)$.
}
\label{tbl:films}
\vskip3mm
{
\setlength{\tabcolsep}{2.8pt}
\begin{tabular}{|c|c|c|c|c|c|c|c|}
\hline
Mat. & Ref. & $T_{c0}$, K & $d$, nm & $l$, \AA & $\gamma^{-1}_\text{fit}$ & ${\cal L}$ & ${\cal L}_d$ \\
\hline
\hline
NbN & \cite{NoatRodichev_NbN_2013} & 15 & $2\div15$ & $\sim5$ & 5.0 & 5.7 & $5.6\div 3.4$ \\
\hline
NbN & \cite{Makise_NbN_2015} & 15 & $1\div26$ & 2 & 8.3 & 7.2 & $6.2\div2.1$\\
\hline
NbN & \cite{ChandRaychaudhuri_NbN_2012} & 17 & $>50$ & $<7$ & $-$ & 4.8 & 3D \\
\hline
TiN & \cite{SacepeBaturina_TiN_2008} & 5 & $3.6\div5$ & 3 & 6.2 & 8.9 & $6.4\div2.4$ \\
\hline
MoGe & \cite{GraybealBeasley_MoGe_1984,Finkelstein_TcSCfilms_JETP_1987} & 7 & $1.5\div100$ & $\sim4$ & 8.2 & 6 & $<4.0$ \\
\hline
MoSi & \cite{Fogel_MoSi_1996} & 7 & $1\div20$ & 5 & 7.0 & 5.6 & $<4.7$ \\
\hline
MoC & \cite{SzaboSamuely_MoC_2016} & 8 & $3\div30$ & $<4$ & 7.5 & 5.5 & $3.2\div0.9$ \\
\hline
WRe & \cite{Raffy_WRe_1983} & 6 & $3\div120$ & 4 & 7.4 & 6.1 & $<2.7$ \\
\hline
Nb & \cite{Couedo_Nb(Si)_2012} & 7 & $2.5\div26$ & 18 & 11.7 & 5.2 & $<4.8$ \\
\hline
\end{tabular}%
}
\end{table}

Another problem with interpreting experimental data in terms of Eq.\  \eqref{Fin} is an implicit assumption that the effect of $T_c$ suppression is determined by the dimensionless film conductance only.
However in real thin films, the impurity concentration and hence the mean free path $l$ do vary with the film thickness due to peculiarities of the fabrication process. 
Large amount of experimental data on the critical temperature of thin films has been analysed in Ref.~\cite{IvryBerggren_review_UniversalTcScaling_2014}, where it has been demonstrated that $T_c$ is primarily dependent on the 3D bulk conductivity $\sigma\propto k_F^2l$ rather than the 2D sheet conductance $g\propto k_F^2ld$.

Inapplicability of Eq.\ \eqref{Fin} for the description of $T_c$ suppression in thin films is actually a consequence of (i) too narrow interval of 2D diffusion (from $d$ to $\xi_0$), which appears to be insufficient to explain the observed magnitude of the effect and (ii) the smallness of the prefactor $1/g \sim (k_F l)^{-1} (k_F d)^{-1}$. Hence for a quantitative description of experimental data, one has to specify another mechanism of disorder-induced enhancement of the Coulomb interaction that is not related to 2D diffusion. 

In the present paper, we demonstrate that existing experimental data on $T_c$ suppression in thin films can be convincingly explained assuming that the main contribution stems from the processes of \emph{three-dimensional ballistic} motion of electrons with a typical distance between the interaction point and the point of impurity scattering of the order of several wavelengths. 
Our main result is the amendment of the perturbative expression \eqref{Ovchinnikov_and_Japanese_result} for $T_c$ shift:
\be
\label{result}
\frac{\delta T_c}{T_{c0}}
=
- \frac{\alpha}{k_Fl}
- \frac{\lambda}{ 3 \pi g} \log^3\frac{\hbar}{T_{c0}\tau_d} ,
\ee
where the added first term accounts for the contribution of the 3D ballistic region. We emphasize that since all scales starting from the Fermi wavelengths contribute to $T_c$ suppression, keeping the last term originating from the 2D diffusion region on the background of the first one may be justified only for materials with exceptionally low $T_c$ or very small thickness (in particular, for atomically thin films \cite{Pb_on_Si}).

The coefficient $\alpha$ in Eq.~\eqref{result} is nonuniversal and depends on the details of the interaction and the structure of the random potential. In the model of weak short-ranged electron repulsion (amplitude $\lambda$) and Gaussian white-noise random potential, it is given by 
\be
\label{alpha}
\alpha
=
\frac{\pi \lambda \log^{2}\omega_{D}/T_c}{2( 1 + \lambda \log E_{F}/\omega_D)^2 } .
\ee
For realistic superconducting films with the Coulomb inter\-action one should expect a material dependent value $\alpha\sim1$.

\textbf{2. The model.}
We consider a model of $s$-wave superconductivity with a phonon-mediated electron attraction described by the potential
$V_\text{ph}(\mathbf{r}) = - (\lambdaph/\nu) \delta (\mathbf{r})$
effective in the in the energy strip of $\omega_D$ near the Fermi energy, and a short-range repulsion with the potential $V(\mathbf{r}) = (\lambda/\nu) \delta (\mathbf{r})$ and an energy cutoff at $E_F$. 
We will work in the weak-coupling approximation, $\lambdaph, \lambda \ll 1$, and neglect disorder-induced renormalization of the phonon vertex beyond the ladder approximation
\cite{KeckSchmid_SuperconductvityElectronPhononImpureMetals_1976}.
Disorder is modeled by a random potential with the Gaussian white-noise statistics described by the correlator 
$\langle U(\br) U(\br') \rangle = \delta ( \br - \br' )/2 \pi \nu \tau$, where $\nu$ is the density of states at the Fermi level (for one spin projection) and $\tau$ is the elastic scattering time. 

In the absence of disorder-induced renormalization of the interaction vertices, $T_c$ is given by the standard expression of the Bardeen-Cooper-Schrieffer (BCS) theory:
\be
\label{T_c_BCS}
T_{c0} = \omega_D \exp \left( - 1 / \lambdaBCS \right),
\ee
where the effective coupling constant is
\be 
\label{pseudopotential}
\lambdaBCS 
= 
\lambdaph 
- \frac{\lambda}{1 + \lambda \log E_F/\omega_D} .
\ee
The second term, known as the Tolmachev logarithm in Russia and as the Coulomb pseudopotential in the West, describes the effect of the Coulomb repulsion to the Cooper channel undergoing logarithmic renormalization in the energy window from $E_F$ to $\omega_D$
\cite{BogoliubovTolmachevShirkov_NewMethodSC_1959, MorelAnderson_SCRetardedElectronPhonon_1962, McMillan_TcStrongCoupledSC_1968, SM}.  

The critical temperature is determined by the pole of the Cooper ladder at zero momentum and frequency in the Matsubara diagrammatic technique. 
In the presence of a random potential, the diagrammatic series should be averaged over disorder in every possible way.
In the leading order (no-crossing approximation), this process reduces to independent averaging of the product of the two Green functions, $G_{E} G_{-E}$, connecting the interaction vertices ($\lambdaph$ or $\lambda$), which is done via insertion of a cooperon.
According to the Anderson theorem \cite{Anderson_DirtySuperconductors_1959, AG0a, AG0b}, 
the result is disorder independent and leads to the expressions \eqref{T_c_BCS} and \eqref{pseudopotential} for the critical temperature. 

\begin{figure}
\includegraphics[width=1\linewidth]{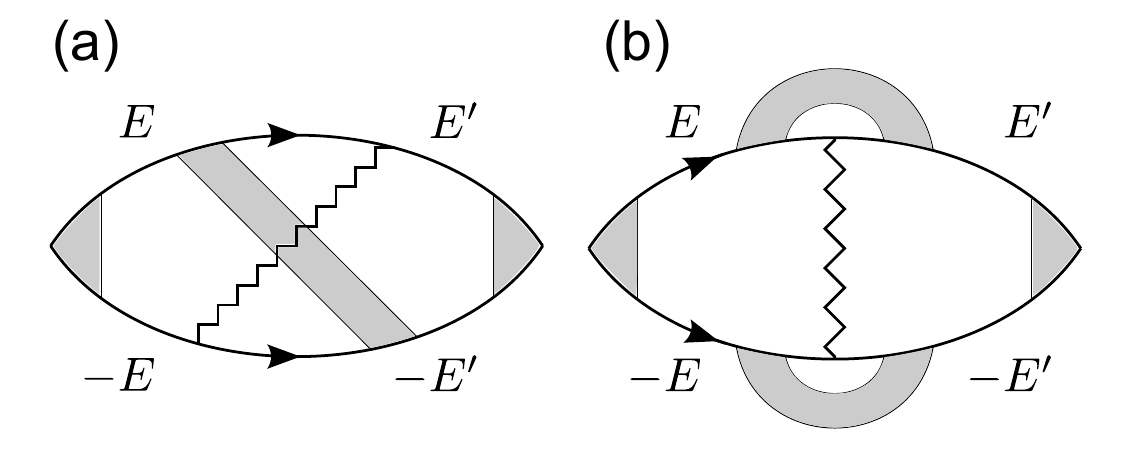}
\protect\caption{
Inelastic diagrams for the diffusive contribution ($q \ll 1/l$ and $E,E' \ll 1/\tau$, where $q$ is the momentum carried by the interaction line) to the Cooper susceptibility that determine $T_c$ shift. The shaded blocks in the center of the diagrams are cooperons and diffusons connecting the Green functions with the opposite Matsubara energy signs. 
The shaded triangles in the corners of the diagrams designate renormalization of the phonon vertex by the impurity ladders and ladders of electron interaction with the constant $\lambda$.} 
\label{fig:Diffusive}
\end{figure}

\textbf{3. Diffusive contribution.} 
In order to find the shift of $T_c$, one has to take into account processes describing an interplay of interaction and disorder in the next order with respect to no-crossing diagrams
\cite{
MaekawaFukuyama_Localization2DSC_1982, 
Maekawa_UpperCriticalField2DSC_1983, EbisawaFukuyamaMaekawa_SCTcDirtyFilms_1985, TakagiKuroda_AndersonLocSCTransition2D_1982,
Finkelstein_TcSCfilms_JETP_1987, Finkelstein_review_SCSuppressionDisorderedFilms_1994}. 
The leading diagrammatic contributions in the diffusive region are shown in Fig.~\ref{fig:Diffusive}, 
where the interaction (zigzag line) is crossed by the impurity ladders --- diffusons and cooperons --- depicted as gray blocks.
The diagram (a) has a mirror counterpart, while the diagram (b) contains two additional contributions with an impurity line connecting the Green functions with the energy of the same sign (Hikami box) \cite{HikamiBox}.
Analytical expression for $T_c$ shift contains a summation over two Matsubara energies $E$ and $E'$:
\be 
\label{dTc_through_I}
\frac{\delta T_{c}}{T_{c0}} 
= 
-\frac{2 \pi\lambda}{\nu} \left( \frac{\lambda_{\text{ph}}}{\lambda_{\text{BCS}}} \right)^2
T^{2}\sum_{E,E'>0}^{E_F}
\frac{u(E) u(E') I_{E,E'}}{ EE'},
\ee
where the factor $\lambda_{\text{ph}} / \lambda_{\text{BCS}}$ and the logarithmic function $u(E) = \theta(\omega_D - E) - ( \lambda \log \omega_D / T ) / (1 + \lambda \log E_F / T)$ represent
renormalization effects, which can be introduced by adding $\lambda$-interaction ladders to the left and right vertex of the diagram \cite{SM}. 
In the diffusive region, the quantity $I_{E,E'}$ in the film geometry can be expressed via an integral over the 2D in-plane momentum $\qxy$ and a sum over the transverse modes of the Laplace operator with the Neumann boundary conditions  
($q_z = 2\pi m/d$, with $m=0$, 1, \dots) carried by the interaction line \cite{SM}:
\be \label{I_general}
I_{E,E'} = 
\frac{\tau}{d} \sum_{q_z}
\int \frac{d \bm q_\parallel}{\left(2\pi\right)^2}
\frac{ f_{q}(E+E')^{2} \left[ 3 - f_{q}(E+E')\right]}{1-f_{q}(E+E')} .
\ee
To trace the crossover to the ballistic region, we write cooperons and diffusons beyond the diffusive approximation and express them via the function
$ f_q(\omega) = (ql)^{-1} \mathop{\rm arctan} [ ql / (1 + |\omega|\tau) ] $,
which corresponds to the one step of the impurity ladder at arbitrary values of $ql$ and $\omega\tau$, but under the conditions $q\ll k_F$, $\omega\ll E_F$.
An analogous approach was used in Ref.~\cite{StepSkv2018} to  calculate the fluctuation conductivity at arbitrary disorder strength. 

\begin{figure}
\includegraphics[width=0.85\linewidth]{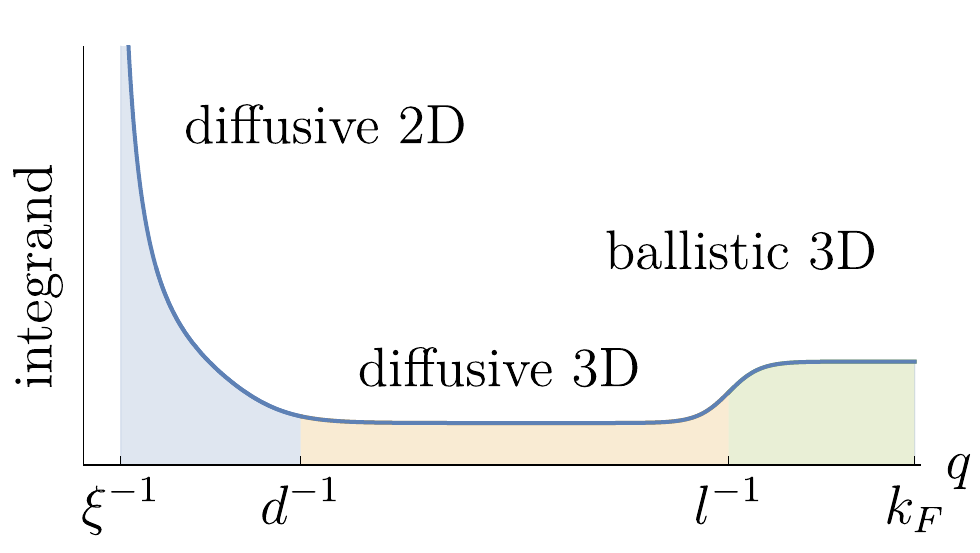}
\protect\caption{
Sketch of the dependence of integrand in Eq.~\eqref{I_general} on $q$ (at not too large $E + E'$). 
In the region $q>1/d$ it has a weak $q$ dependence, changing by a factor of $\pi^2/8$ at the crossover from the diffusive to ballistic motion (at $q\sim 1/l$).
} 
\label{fig:integrand}
\end{figure}

The leading 2D diffusive contribution stems from the mode with $q_z=0$. 
Cutting the integral over $q$ at the momentum $1/d$ and the energy summation at $\omega_D$, and taking into account that for realistic films studied in experiments the Debye frequency $ \omega_D$ is comparable to $\hbar/\tau_d$ \cite{Brun_NbN_2020}, we arrive at the well-known result \eqref{Ovchinnikov_and_Japanese_result} with $\tau_*\sim\tau_d$.
Note however that the extraction of the 2D diffusive contribution out of expressions \eqref{dTc_through_I} and \eqref{I_general} is complicated by the fact that the contributions of other regions are in fact larger. 
Indeed, at the scale $q \sim 1/d$ the 2D logarithmic behavior is changed to a linearly divergent one due to excitation of higher transverse modes, making the momentum integral three-dimensional. 
One can estimate the contribution of the 3D diffusive region by introducing an artificial cutoff at $q \sim 1/l$, which gives
\begin{equation} 
\label{kFl_squared_contribution}
\frac{\delta T_{c}^{\text{(diff, 3D)}}}{T_{c0}} \sim
-\frac{ \lambda}{(k_{F}l)^{2}} \log^{2}\frac{\omega_{D}}{T_{c0}}.
\end{equation}
This contribution has only two out of three logarithmic factors but nevertheless it exceeds Eq.\ \eqref{Ovchinnikov_and_Japanese_result} by the parameter $d/l\gg1$. 
However, nothing prevents considering even greater momenta in Eq.~\eqref{I_general} and study the ballistic region $q\gg1/l$. 
Remarkably, in this region the integrand of Eq.~\eqref{I_general} still obeys the $1/q^2$ behavior, but with a different numerical prefactor. 
This means that the main contribution to the integral originates from momenta of the order of Fermi momentum, $q \sim k_F$. 
This region requires a special treatment, which will be done below.
Schematically the role of different momentum regions is illustrated in Fig.~\ref{fig:integrand}. 
Up to logarithmic factors coming from the energy summations, the integral of the shown curve determines the contribution of the corresponding regions to $T_c$ shift. 

\textbf{4. Ballistic contribution.}
In this Section, we study the ballistic contribution to $T_c$ shift originating from processes with momentum transfer $q > 1/l$.
Due to the assumption $l\ll d$, electron motion can be assumed to be three-dimensional.
This contribution is described by the diagrams shown in Fig.~\ref{fig:Diffusive}, 
where we left only one impurity line out of the diffusive ladder, corresponding to scattering on one impurity. 
For an accurate calculation, one should reconsider expression \eqref{I_general}, relaxing the assumption $q\ll k_F$.

\begin{figure}
\includegraphics[width=0.95\linewidth]{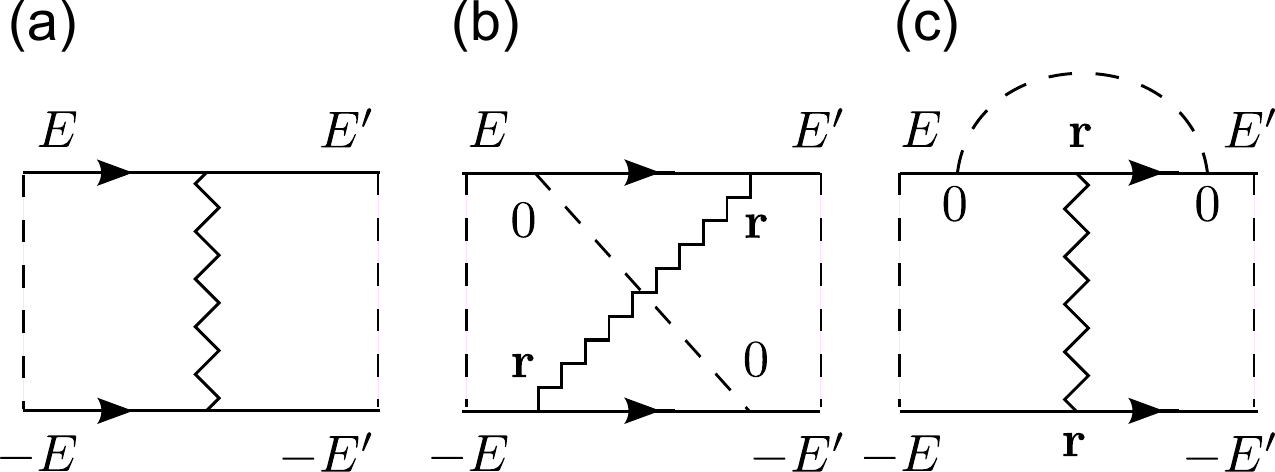}
\protect\caption{(a) Electron-electron interaction vertex $\lambda$ in the Cooper channel with the first impurity line of the surrounding cooperons. (b), (c)
Diagrams describing the leading vertex correction $\lambda^c$ from the ballistic region. Both diagrams have mirrored counterparts.}
\label{fig:Ballistic}
\end{figure}

The ballistic contribution can be described as a correction to the bare (unrenormalized) repulsive electron-electron coupling constant in the Cooper channel, $\lambda^c$, which in the leading order coincides with 
 $\lambda$ (Fig.~\ref{fig:Ballistic}(a)).
The leading corrections are given by the diagrams Fig.~\ref{fig:Ballistic}(b) and Fig.~\ref{fig:Ballistic}(c).
In the considered model of point-like interaction and delta-correlated disorder, the calculation of these diagrams can be performed analytically and leads, generally speaking, to an energy-dependent correction $\delta \lambda_{EE'}^c$ to the Cooper-channel coupling:
\be \label{delta_lambda_through_P}
\frac{\delta\lambda_{E,E'}^c} {\lambda} = 2 \frac{(\text{b}) + (\text{c})}{(\text{a})} 
=
\frac{2[ P(E,E') + P(E,-E') ]}{(2\pi\nu\tau)^2 f_0(2E)f_0(2E') \lambda} ,
\ee
where the terms in the brackets correspond to the diagrams (b) and (c), respectively, and the numerical coefficient $2$ is due to 
mirrored diagrams.
The factors $f_0(\omega)=1/(1+|\omega|\tau)$ in the denominator originate from the momentum integration of a pair of the Green functions in Fig.~\ref{fig:Ballistic}(a) (one step of the diffusive ladder).

It is convenient to calculate block $P(E,E')$ in the coordinate representation \cite{AntSkv2020}. 
Since the electron-electron interaction as well as the disorder correlator are assumed to be point-like, analytical expression contains only one integral over the distance $\bm{r}$ between the impurity and the interaction point, so we get:
\be 
\label{P_in_r}
P(E,E') = \frac{\lambda}{2 \pi \nu \tau} \int d\bm{r}\, G_{+}G'_{-} [G_{+} G_{-}] [G'_{+} G'_{-}],
\ee
where $G_{\pm} = G_{\pm E}(\bm{r})$ 
are disorder-averaged Green functions and the prime refers to the energy argument $E'$.
The square brackets denote the real-space convolution: $\left[G_{+} G_{-} \right] = \int G_+(\bm{\rho}) G_-(\bm{r}-\bm{\rho}) \, d \bm{\rho} $.
As will be demonstrated below, the integral over $\bm{r}$ in Eq.~\eqref{P_in_r} converges on the scale $1/k_F$ that allows to replace the Green functions by their values in the absence of disorder:
\be
\label{G_GG}
G_\pm = - \pi\nu\frac{e^{\pm ik_{F}r}}{k_{F}r} ,
\quad\!
\left[G_{+} G_{-}\right] = \frac{2\pi\nu\tau}{1 + 2|E|\tau} \frac{\sin k_{F}r}{k_{F}r} ,
\ee
where the convolution was calculated under the assumption $E,E' \ll E_F$.

One can easily show that the integral in Eq.~\eqref{P_in_r} vanishes for different signs of the energies $E$ and $E'$, and thus $P(E,E') \propto \theta(E E')$. 
Thereby in the considered model, the ballistic diagrams in Figs.~\ref{fig:Ballistic}(b) and \ref{fig:Ballistic}(c)
are nonzero for the same relation between the energy signs as for the diffusive diagrams in Figs.\ref{fig:Diffusive}(a) and \ref{fig:Diffusive}(b), respectively. 
This conclusion is \emph{a priori} not obvious because a single impurity line can connect two Green functions of the same energy sign.
However, we see that in the case of the point-like interaction and delta-correlated disorder, these diagrams vanish in the ballistic limit as well. 

Substituting Eq.\ \eqref{G_GG} to Eq.\ \eqref{P_in_r} and then to Eq.~\eqref{delta_lambda_through_P}, 
we observe that the factors $(1 + 2 |E| \tau)$ and $(1 + 2 |E'| \tau)$ in the denominators of $[G_+ G_-]$ and $[G_+' G_-']$ cancel the same factors $f_0(E)$ and $f_0(E')$ in Eq.~\eqref{delta_lambda_through_P}. 
The only energy dependence of $\delta \lambda_{E,E'}^c$ is thus due to the factor $\theta(EE')$ contained in the block $P(E,E')$. 
However, it also disappears because of the structure of Eq.\ \eqref{delta_lambda_through_P}.
As a result, the correction $\delta \lambda_{E,E'}^c$ appears to be energy-independent:
\be \label{int_over_r}
\delta\lambda^c
=
\frac{\pi\nu\lambda}{2\tau}
\int \frac{d\bm{r}}{(k_{F}r)^{2}}
\left(\frac{\sin k_{F}r}{k_{F}r}\right)^{2} = \frac{\pi\lambda}{2k_F l} .
\ee
As expected, the integral stems from the scales of the order of the electron wavelength, which is typical for 3D mesoscopic effects
\cite{BelitzKirkpatrick_UltrasonicAttenuationDisordered_1986, SkipetrovTiggeln_DOSFluctuationsSpeckleCorrelations_2006, SmolyarenkoAltshuler_RareEvents_1997}.

\begin{figure*}
\includegraphics[width=0.64\columnwidth]{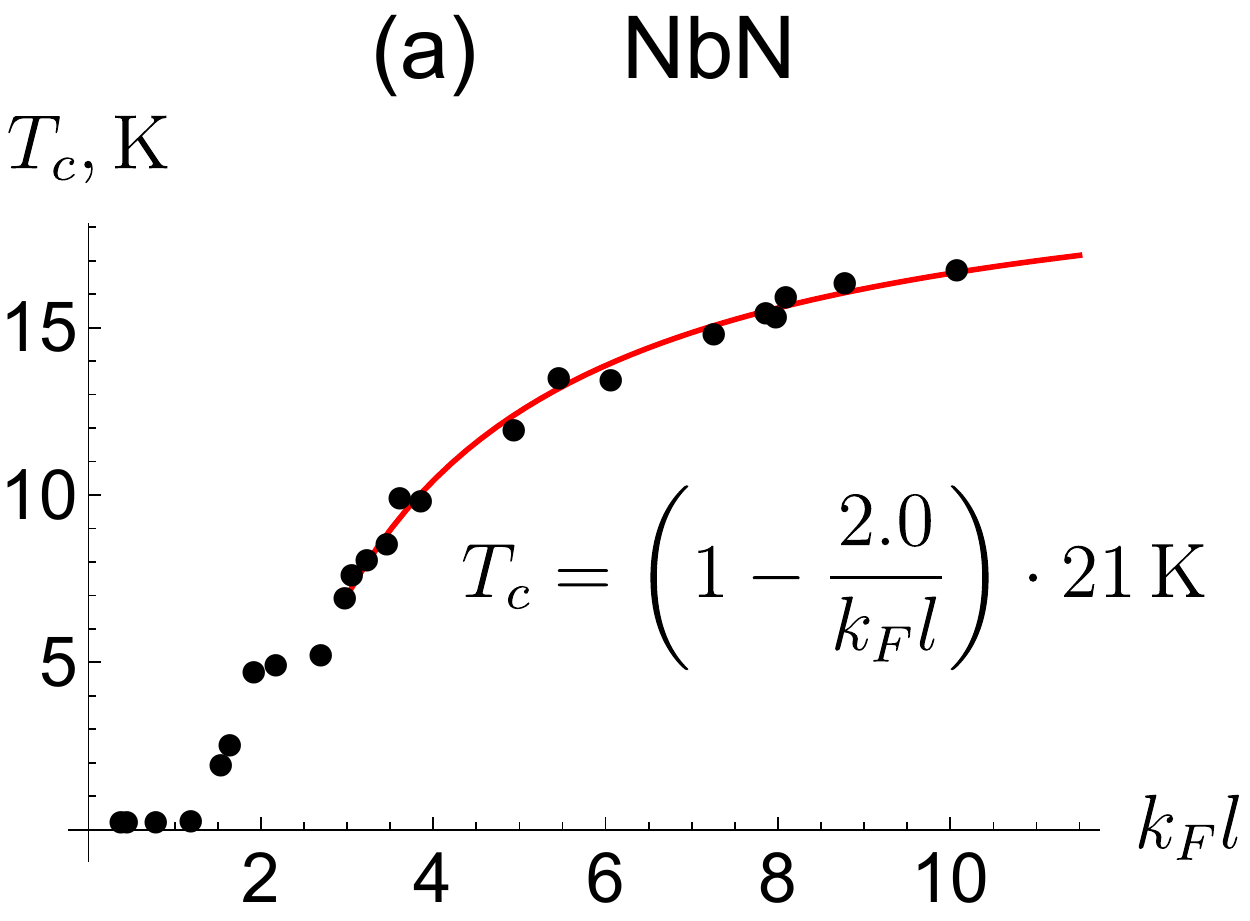}
\hspace{2mm}
\includegraphics[width=0.64\columnwidth]{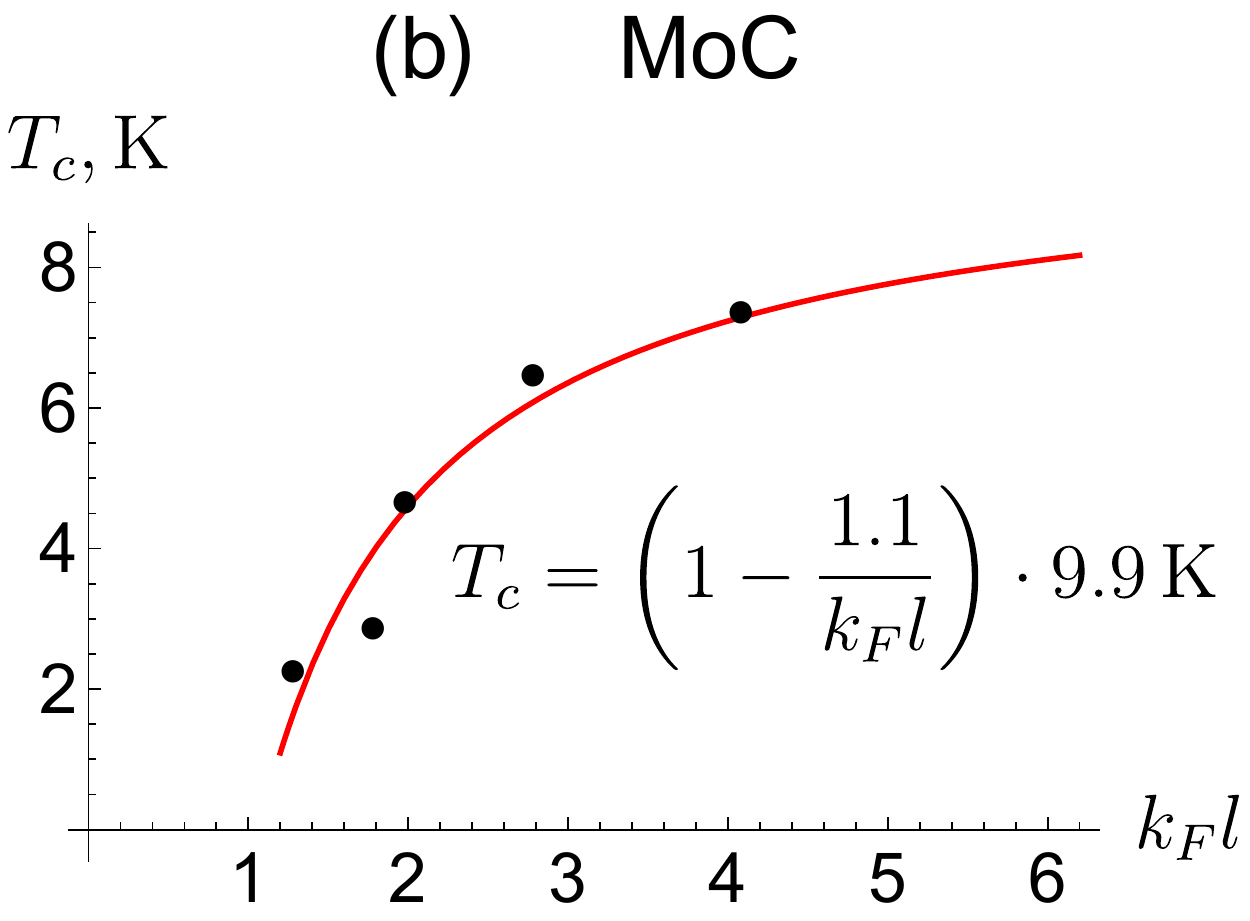}
\hspace{2mm}
\includegraphics[width=0.64\columnwidth]{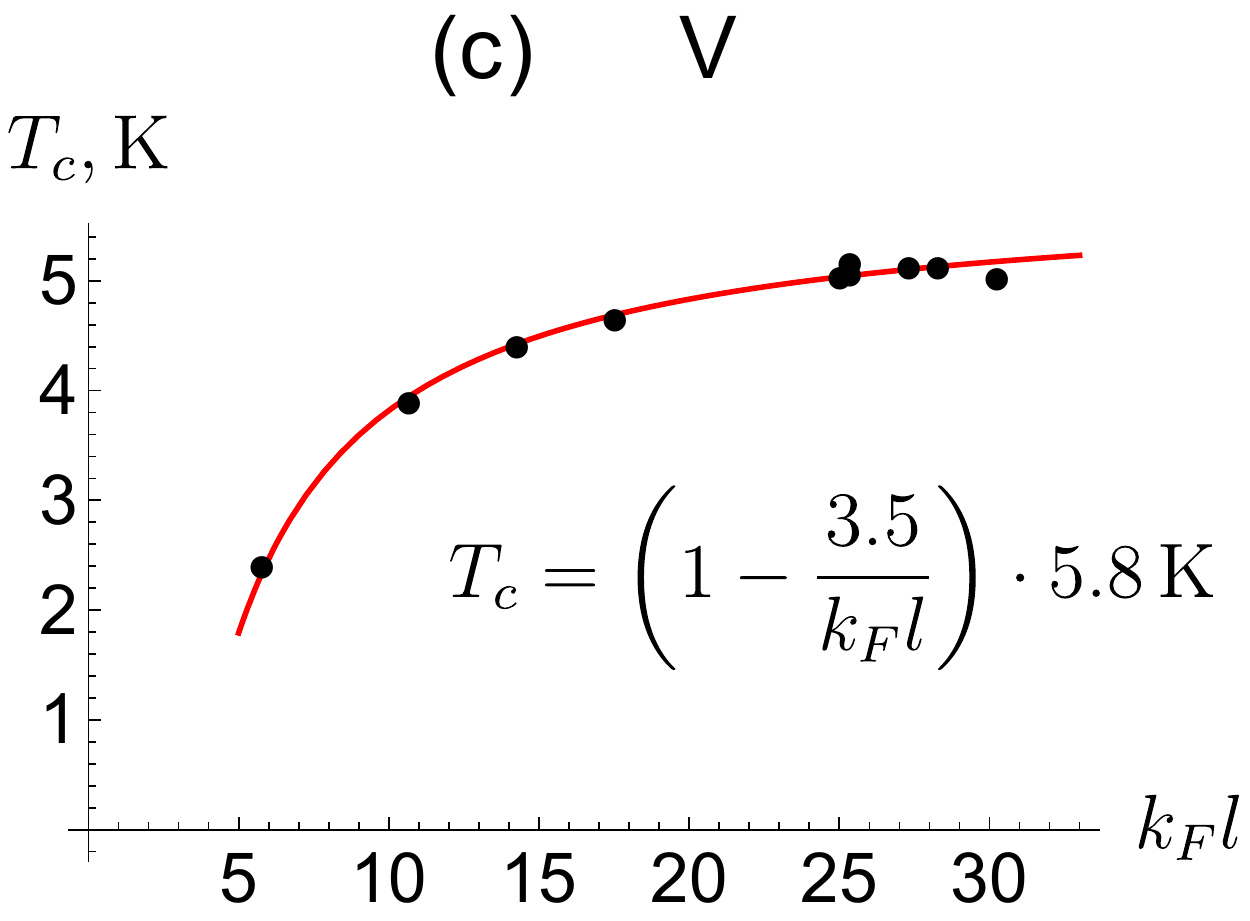}
\caption{
Experimental data on the dependence of $T_c$ on $k_F l$ (dots) and their fitting with the help of Eq.\ \eqref{Tc-alpha} (solid line) for superconducting films of different thickness and composition:
(a) NbN \cite{ChandRaychaudhuri_NbN_2012}, (b) MoC \cite{SzaboSamuely_MoC_2016}, (c) V \cite{Teplov_V_1976}.}
\label{fig:exper_fit}
\end{figure*}

The obtained correction may be interpreted in the spirit of Ref.~\cite{Glazman} as the renormalization of the contribution of electron-electron interaction to the Cooper channel due to scattering on Friedel oscillations caused by impurities.
This correction describes the enhancement of the electron-electron repulsion, leading to the increase of the Coulomb pseudopotential and, consequently, to the suppression of the effective coupling constant
 $\lambdaBCS$. 
Suppression of $T_c$ can be found by substituting $\lambda$ by $\lambda+\delta\lambda^c$ 
 and expanding Eq.~\eqref{pseudopotential} in $\delta\lambda^c$:
\be \label{main_part:ballistic_contr}
\frac{\delta T_{c}^{(\text{ball, 3D})}}{T_{c0}} 
= 
- \frac{\pi}{2}
\frac{\lambda}{k_F l} 
\left( \frac{ \log \omega_D/T_{c0} } {1 + \lambda \log E_F/\omega_D} \right)^2.
\ee

\textbf{5. Role of elastic diagrams.}
Besides \textit{inelastic} diagrams shown in Figs.~\ref{fig:Diffusive} and \ref{fig:Ballistic}, 
where the interaction line connects the upper and lower Green functions, there is a set of \textit{elastic} diagrams related to the interaction correction to the one-particle Green function.
As demonstrated by Finkel'stein \cite{Finkelstein_review_SCSuppressionDisorderedFilms_1994} for 2D diffusion, the contribution of this set of diagrams is always small: at $Dq^2>\omega$ they contain a smaller power of a large logarithm, while at $Dq^2<\omega$ their contribution is canceled by contributions of inelastic diagrams and of an additional set of diagrams restoring the gauge invariance of the theory. 
The latter diagrams become subleading already in the diffusive region at $Dq^2 > \omega$ and therefore are not considered in the present paper. 

In the case of an instantaneous electron-electron interaction, there is an exact relation  \cite{MaekawaFukuyama_Localization2DSC_1982,
Maekawa_UpperCriticalField2DSC_1983, Belitz-correlation-gap} 
between the contribution of elastic diagrams to $T_c$ shift and correction to the tunneling density of states $\delta\nu(\varepsilon)$, which can be represented \cite{SM} in the form analogous to Eq.~\eqref{dTc_through_I}:
\be
\label{elastic_via_DOS}
\frac{\delta T_{c}^{\text{(elast)}}}{T_c} 
= 
\left( \frac{\lambda_{\text{ph}}}{\lambda_{\text{BCS}}} \right)^2 
T \sum_{E} \int d \varepsilon \, \frac{u^2(E)}{E^2 + \varepsilon^2} \frac{\delta \nu(\varepsilon)}{\nu_0} .
\ee
We will use known results for $\delta\nu(\epsilon)$ in order to estimate the contribution \eqref{elastic_via_DOS} of elastic diagrams. 

The correction to the tunneling density of states of a 3D metal in the diffusive region ($|\varepsilon| < 1 / \tau$) has the form $\delta \nu_\text{diff} (\varepsilon) / \nu_0 \sim \lambda \sqrt{|\varepsilon| \tau} / (k_F l)^2$ \cite{AltshulerAronov_DOS_ZBA_1979}. 
A simple algebra reveals that the contribution to $T_c$ shift from this region is proportional to $1/(k_F l)^2$, which is parametrically smaller than the contribution of the ballistic region discussed below. 

The correction to the tunneling density of states in the 3D ballistic region ($|\varepsilon| > 1/\tau$) was studied in Refs.\ \cite{Koulakov_QuasiballisticDOS_2000, AntSkv2020} and appeared to be linear in energy and generally asymmetric with respect to the Fermi level. 
In the case of a point-like interaction, delta-correlated disorder, and parabolic electron spectrum, it is finite only for energies below the Fermi energy and has the form $\delta \nu_\text{ball} (\varepsilon) / \nu_0 \sim \lambda |\varepsilon| \theta(-\varepsilon) / (k_F l)$ \cite{AntSkv2020}. 
Then Eq.~\eqref{elastic_via_DOS} yields
\be 
\frac{\delta T_{c}^{(\text{ball, 3D, elast}) } }{ T_{c0} } 
\sim
\frac{\lambda^3}{k_F l} 
\left( \frac{ \log \omega_D/T_{c0} } {1 + \lambda \log E_F/\omega_D} \right)^2,
\ee 
which is parametrically smaller than the leading contribution \eqref{main_part:ballistic_contr} under the model assumption $\lambda \ll 1$.
The absence of a linear-in-$\lambda$ contribution from elastic diagrams is related to the fact that, contrary to Eq.\ \eqref{dTc_through_I} with two logarithmic summations over $E$ and $E'$, the integral \eqref{elastic_via_DOS} in the 3D ballistic region is not logarithmic.
The conclusion that elastic diagrams do not contribute to the leading $T_c$ shift is presumably quite general and related to the fact that the tunneling density of states is not a thermodynamic quantity. 

\textbf{6. Conclusion.}
In the present paper we studied the influence of the 3D ballistic region of electrons motion on the critical temperature degradation of moderately disordered superconducting films ($k_Fl\gg1$). 
Assuming the model of a point-like repulsion and delta-correlated disorder, we calculated the perturbative contribution of this region to $T_c$ suppression given by the first term in Eq.~\eqref{result}. 
When comparing our theory with experimental data, one should take into account that in real samples $\lambda\sim1/2$ due to the Coulomb interaction and that the numerical factor in Eq.~\eqref{alpha} is model-specific.  
In general, one might expect that the ballistic contribution to $T_c$ shift has form $\delta T_c/T_{c0} = -\alpha/k_Fl$ 
with $\alpha\sim1$. 

The second term in Eq.~\eqref{result} describes the standard contribution to $T_c$ suppression originating from the region of two-dimensional electron diffusion, where the logarithm stems from the spatial scales between the film thickness $d$ and the coherence length $\xi_0$.
The smallness of this interval for realistic films and a relatively large value of the dimensionless conductance $g\sim(k_Fl)(k_Fd)$ makes it practically negligible compared to the three-dimensional ballistic contribution.

Fig.~\ref{fig:exper_fit} presents the fits of experimental data ($T_c$, $k_Fl$) for superconducting films of different thicknesses made of three different materials following the fermionic scenario of $T_c$ suppression by the formula
\be
\label{Tc-alpha}
T_c = (1-\alpha/k_Fl)T_{c0},
\ee 
where $\alpha$ and $T_{c0}$ are treated as fitting parameters.
A rather good agreement is observed, with the material-dependent value of  $\alpha$ being of the order of one, as expected. 
We emphasize that the data for NbN presented in Fig.~\ref{fig:exper_fit}(a) refer to thick films \cite{ChandRaychaudhuri_NbN_2012}, for which there is no two-dimensional diffusive region at all (see Table \ref{tbl:films}).

Based on (i) the observed agreement between experimental data and Eq.\ \eqref{Tc-alpha}, (ii) intrinsic inconsistencies of the theory behind Eq.~\eqref{Fin} mentioned above, and (iii) the findings of Ref.~\cite{IvryBerggren_review_UniversalTcScaling_2014}, which indicate that $T_c$ is primarily dependent on the 3D conductivity rather than the 2D sheet conductance, we make the following practically relevant conclusion:

\emph{
For a substantial fraction of not too thin moderately disordered superconducting films that follow the fermionic scenario of superconductivity suppression, the latter is governed by the proximity to the threshold of three-dimensional Anderson localization and controlled by the parameter $k_Fl$.
Two-dimensional diffusion effects, controlled by dimensionless conductance $g$ are also present, but they typically constitute only a small correction on top of three-dimensional ballistic effects. 
}

The authors are grateful to I. S. Burmistrov, M.~V.~Feigel'man, A. M. Finkel'stein, P. Samuely, P.~Szab\'o, K. S. Tikhonov, and P. M. Ostrovsky for useful discussions. The research was supported by a grant from the Russian Science Foundation No.~20-12-00361.

\bibliography{JETPLetters_arxiv}

\begin{thebibliography}{72}
\bibitem [{com()}]{com-table}%
  \BibitemOpen
  \href@noop {} {}\bibinfo {note} {For WRe and TiN films, we took $1/k_F \sim l
  \sim a$, where $a$ is the interatomic distance. For MoC films, the free
  electron mass was taken for the effective mass.}\BibitemShut {Stop}%
\end{thebibliography}


\clearpage
\onecolumngrid
\section{SUPPLEMENTAL MATERIAL\\[6pt]
}
\renewcommand{\theequation}{S\arabic{equation}}
\renewcommand{\thefigure}{S\arabic{figure}}
\setcounter{equation}{0}
\setcounter{figure}{0}

\maketitle

\subsection{Cooper susceptibility and the critical temperature}

The starting point of our analysis is the zero-momentum Cooper susceptibility: 
\be \label{Pi_full}
	L = \int d\bm{r} \int^{1/T}d\tau\,
  \bigl\langle \psi_{\downarrow}^{+}(\bm{r},\tau)\psi_{\uparrow}^{+}(\bm{r},\tau)\psi_{\uparrow}(0,0)\psi_{\downarrow}(0,0)\bigr\rangle. 
\ee
The divergence of $L$ as a function of temperature $T$ marks the transition to the superconducting state. 

The basic element of the theory is the disorder-averaged Matsubara Green function
\be
	G_{\pm E} (\mathbf{k}) = \frac{1}{\pm iE - \xi_{\mathbf{k}} \pm i/2 \tau} .
\ee
For calculations in the momentum representation, we use the approximation $\xi_{\mathbf{k}} = \upsilon_F  \left( |\mathbf{k}| - k_F \right)$, which breaks down in the vicinity of the Fermi momentum. When working in the real space, we assume a parabolic dispersion of the electron spectrum: $\xi_{\mathbf{k}} = k^2 / 2 m - E_F$. 

In order to calculate $L$, we need to draw all possible diagrams with the interaction vertices $\lambdaph$ and $\lambda$, and average them over disorder. It is convenient to calculate ladders of repulsive interaction lines $\lambda$ first and then insert the corresponding block (denoted as $\Pi$) between the attractive phonon lines $\lambdaph$. Summing the corresponding ladder, we obtain
\be
\label{L_via_Pi}
	L  = \frac{\Pi}{1 - \lambdaph \Pi / \nu} 
\ee
As the block $\Pi$ is inserted between the phonon lines, energy cutoff at the Debye frequency $\omega_D$ is implied at its edges. 
Equation \eqref{L_via_Pi} allows to express the critical temperature in terms of $\Pi$ through the relation
\be
\label{Tc-eq-Pi}
	\nu \lambdaph^{-1} = \Pi(T_c) .
\ee

\subsection{Ballistic disorder ladders}
In the following calculation we will need the expression for the ``ballistic'' cooperon and diffuson $\mathcal{C}(\mathbf{q},\omega)$ derived at arbitrary values of $q l$, $\omega \tau$ (but we still assume that $q \ll k_F$ and $\omega \ll E_F$). Taking $E>0$ and $E-\omega<0$, we get for one step of the ladder \cite{StepSkv2018}: 
\be \label{fq}
	f_{q}(\omega) 
  = \frac{\nu}{2\pi\nu\tau}\int\frac{d\Omega}{4\pi}\int d\xi\,\frac{1}{iE-\xi+{i}/{2\tau}}\frac{1}{i(E-\omega)-\xi-\bm{\upsilon q}-{i}/{2\tau}} 
= \frac{1}{ql}\arctan\frac{ql}{1+\omega\tau}.
\ee
Summing the geometric series of the diffusive ladder, we obtain
\be
\label{ballistic_diffuson}
	\mathcal{C}(\mathbf{q},\omega)=\frac{1}{2\pi\nu\tau}\frac{1}{1-f_{q}\left(\omega\right)}; \qquad \quad \mathcal{C}(0,\omega)=\frac{1}{2\pi\nu\tau}\frac{1+\omega\tau}{\omega\tau}.
\ee

\begin{figure}[h]
	\includegraphics[width=0.8\linewidth]{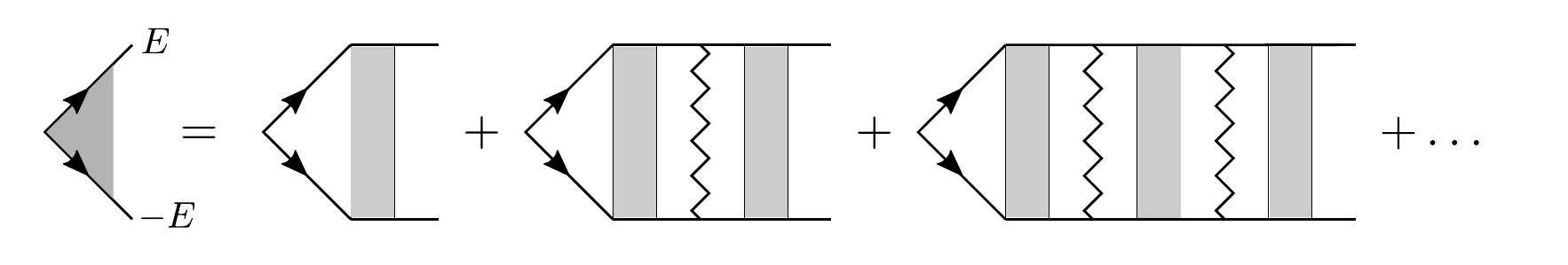}
	\protect\caption{Diagrammatic equation for the renormalized Cooper vertex $\upsilon(E)$. Zigzag lines stand for the repulsive interaction $\lambda$. Blocks with no impurity lines between the interaction lines are also included. The outer Green functions are not included into the expression for $\upsilon(E)$.}
	\label{fig:vertex} 
\end{figure}

\subsection{Renormalization of the phonon vertex}

In order to deal with logarithmic contributions originating from various energy intervals, it is convenient to introduce the renormalized phonon vertex $\upsilon(E)$ defined as the sum of the sequence of diagrams shown in Fig.~\ref{fig:vertex}. 
In brief, $\upsilon(E)$ takes into account ladders of the interaction lines (repulsion constant $\lambda$), which are known to be responsible for the ``Tolmachev logarithm'' (Morel-Anderson pseudopotential) renormalisation \cite{BogoliubovTolmachevShirkov_NewMethodSC_1959, MorelAnderson_SCRetardedElectronPhonon_1962, McMillan_TcStrongCoupledSC_1968}.
In the quasiballistic region it is important to account for the diagrams, where the diffusive ladder may be absent (no impurity lines). 
Since the vertex contains the photon interaction, the energy arguments in the pair of Green functions adjacent to the vertex should be smaller than $\omega_{D}$. This property is taken into account by introducing the step function $\theta(\omega_{D}-|E|)$ to the first term of the series and restricting integrations over internal energies in the other terms (see below). 
The renormalised phonon vertex then takes the form
\be
\label{upsilon}
\upsilon(E) = 
\left[
1 + \frac{2\pi\nu\tau}{1+2|E|\tau}\mathcal{C}(0,2|E|)\right] u(E) 
= \frac{1 + 2 E \tau} {2 E \tau} u(E),
\ee
where
\be
\label{V(E)}
u(E) = \theta(\omega_D - |E|) - \frac{\lambda}{\nu} T \sum_{E'}^{\omega_D} \frac{\pi\nu}{E'} + \left(-\frac{\lambda}{\nu} \right)^{2} T^2 \sum_{E'}^{\omega_D} \frac{\pi\nu}{E'} \sum_{E''}^{E_F} \frac{\pi\nu}{E''} + \dots = \theta(\omega_D - |E|) - \frac{\lambda \log \omega_{D}/T}{1 + \lambda \log E_{F}/T}.
\ee

\begin{figure}[b]
	\includegraphics[width=0.3\linewidth]{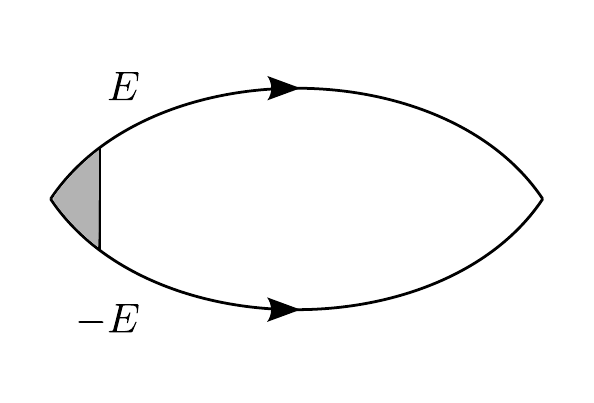}
	\protect\caption{Cooper bubble $\Pi_0$ in the no-crossing approximation.}
	\label{fig:Pi0} 
\end{figure}

\subsection{Anderson theorem}

In the leading no-crossing approximation, $\Pi$ is given by the diagram depicted in the Fig.~\ref{fig:Pi0} and is given by
\be \label{Pi0}
	\Pi_0(T) = 2 \pi \nu \tau \sum_E^{\omega_D} f_0 (2 |E|) \upsilon(E) 
  = \nu \frac{1+\lambda \log E_F/\omega_D}{1+\lambda \log E_F/T} \log \frac{\omega_D}{T}. 
\ee
This expression appears to be disorder-independent, which leads to the insensitivity of the critical temperature to potential disorder in the leading order (Anderson theorem) \cite{Anderson_DirtySuperconductors_1959, AG0a, AG0b}.
Solving Eq.\ \eqref{Tc-eq-Pi} with $\Pi = \Pi_0$, we get the standard Bardeen-Cooper-Schrieffer (BCS) expression \eqref{T_c_BCS} with the renormalised coupling constant $\lambdaBCS$ given by Eq.~\eqref{pseudopotential}.

\subsection{Crossing corrections to $\Pi(T)$}

Contributions to $\Pi$ beyond the non-crossing approximation are responsible for the shift of $T_c$. Assuming that $\delta \Pi$ is small and linearizing, we get the following equation for $\delta T_c$ in the first order:
\be
	\nu \lambdaph^{-1} =  \Pi_0(T_{c0}+\delta T_c) + \delta \Pi (T_{c0}).
\ee
Hence we get for the perturbative shift of $T_c$:
\be \label{dTc_through_Pi}
	\frac{\delta T_c}{T_{c0}}  
  = 
  \frac{\delta \Pi}{\nu} 
  \left( \frac{1 + \lambda \log E_F/T_{c0} }{1 + \lambda \log E_F/\omega_D}\right)^2 
= \frac{\delta \Pi}{\nu} \left( \frac{\lambdaph}{\lambdaBCS}\right)^2 .
\ee
In general, account for the renormalisation effects can be done with the help of Eq.\ \eqref{dTc_through_Pi} and insertion of renormalised Cooper vertices $\upsilon(E)$ into the ends of the diagrams, which describe the correction to the Cooper bubble, $\delta \Pi$. This procedure leads to Eqs.~\eqref{dTc_through_I} and \eqref{elastic_via_DOS}.

Applying this technique to the ballistic vertex correction reproduces the result obtained in the Letter by interpreting this correction as a shift of the bare Cooper-channel constant $\delta \lambda^c$ and expanding Eqs.~\eqref{T_c_BCS} and \eqref{pseudopotential}.
On the other hand, applying the same technique to corrections originating at energies $E < \omega_D$ (as the main part of the 2D diffusive Finkel'stein-Ovchinnikov correction) leads to the cancellation of the renormalisation factors.

\subsection{Momentum-space calculation of the critical temperature shift}

Below we sketch the derivation of Eqs.~\eqref{dTc_through_I} and \eqref{I_general}, which represent the contribution of inelastic diagrams (depicted in Fig.~\ref{fig:Diffusive}) to the $T_c$ shift. The calculation is done in the momentum representation in terms of ballistic diffusons and cooperons [see Eq.~\eqref{ballistic_diffuson}] to assess the crossover to the ballistic region. 

The first diagram in Fig.~\ref{fig:Diffusive} represents a correction $\delta \Pi_a$ to be inserted between the phonon lines in the Cooper ladder. When substituted to Eq.~\eqref{dTc_through_Pi} it results in Eq.~\eqref{dTc_through_I} with 
\be
\label{I_first_diagramm}
	I_{E,E'}^{(a)} = 
\frac{\tau}{d} \sum_{q_z}
\int \frac{d \bm q_\parallel}{\left(2\pi\right)^2}
\frac{ f_{q}(E+E')^{2}} {1-f_{q}(E+E')} ,
\ee
where Eqs.~\eqref{fq} and \eqref{ballistic_diffuson} were used and summation over diffusive modes in the film geometry is implied [to be replaced by usual 3D integration $\int (d^3 \bm{q})$ when studying crossover from 3D diffusion to 3D ballistics]. The numerator in Eq.\ \eqref{I_first_diagramm} represents two triangular Hikami boxes in the diagram, while the denominator corresponds to the Cooperon ladder.

\begin{figure}
	\includegraphics[width=1\linewidth]{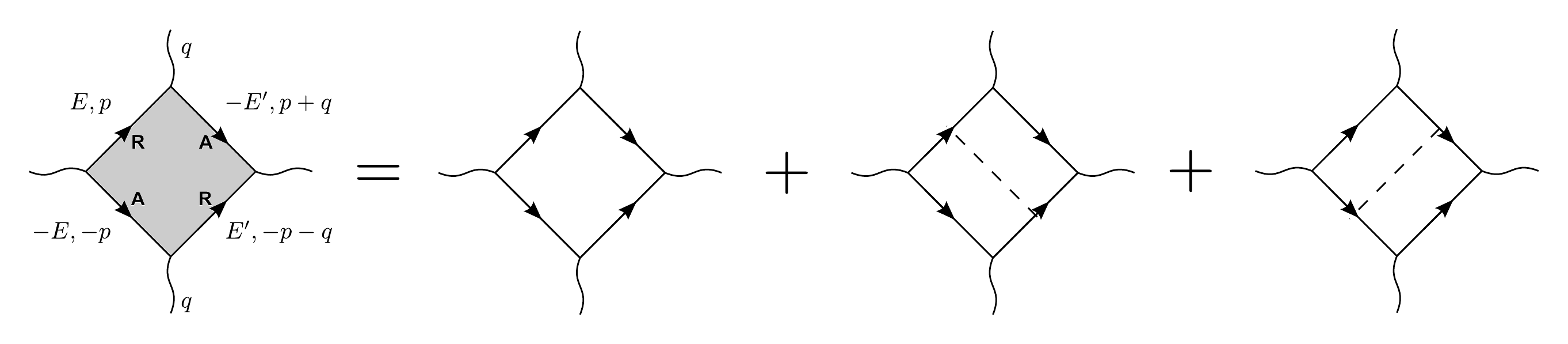}
	\protect\caption{Hikami box $H(q, E, E')$ made of four Green functions.}
	\label{fig:Hikami_box} 
\end{figure}

Calculation of the second diagram in Fig.~\ref{fig:Diffusive} involves computation of the ballistic Hikami box made of of four Green functions (see Fig.~\ref{fig:Hikami_box}), which is given by
\be
\label{Hikami_box}
	H(q, E, E') = 4 \pi \nu \tau^3 \frac{f_q(E+E') [1 - f_q(E+E')]}{(1 + 2 E \tau)(1 + 2 E' \tau)} .
\ee
Then the contribution of the second diagram is given by Eq.~(7) with
\be
\label{I_second_diagramm}
	I_{E,E'}^{(b)} = 
\frac{\tau}{d} \sum_{q_z}
\int \frac{d \bm q_\parallel}{\left(2\pi\right)^2}
\frac{ f_{q}(E+E')^{3} [1-f_{q}(E+E')]} {[1-f_{q}(E+E')]^2} ,
\ee
where the denominator originates from two diffusons in the central part of the diagram and the numerator is the Hikami box \eqref{Hikami_box} multiplied by two additional $f_q (E+E')$ factors, stemming from the integrals of the ``bubbles'' $G_E(p) G_{-E'}(p+q)$ and $G_{-E}(p') G_{E'}(p' - q)$. 

Finally, in order to be able to trace a crossover to the ballistic region, one should also include the diagram obtained from the second diagram in Fig.~\ref{fig:Diffusive} by leaving only one out of the two diffusons encircling the interaction line. That leads to Eq.~\eqref{dTc_through_I} with
\be
\label{I_second_diagramm_prime}
	I_{E,E'}^{(b')} = 
2 \frac{\tau}{d} \sum_{q_z}
\int \frac{d \bm q_\parallel}{\left(2\pi\right)^2}
\frac{ f_{q}(E+E')^{2} [1-f_{q}(E+E')]} {1-f_{q}(E+E')} .
\ee
Finally, summing Eqs.~\eqref{I_first_diagramm}, \eqref{I_second_diagramm}, and \eqref{I_second_diagramm_prime}, one arrives at Eq.~\eqref{I_general}.

\subsection{Elastic diagrams}
\begin{figure}
	\includegraphics[width=0.6\linewidth]{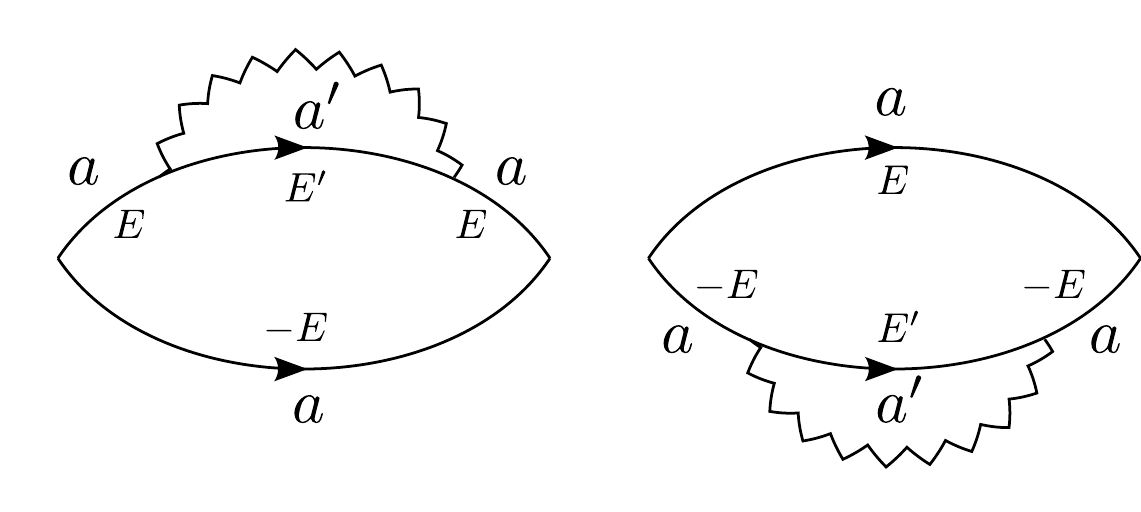}
	\protect\caption{The central part of elastic diagrams to the Cooper susceptibility in the exact-eigenstates representation.}
	\label{fig:elastic_diagrams} 
\end{figure}
The central part $\delta P^\text{elastic}$ of elastic diagrams is depicted in Fig.~\ref{fig:elastic_diagrams}, where we work in terms of the exact eigenstates (labeled by $a, a'$) of the Hamiltonian in the presence of disorder. The corresponding analytical expression is:
\be
\label{P_elastic}
	\delta P^\text{elastic} =  T^2 \sum_{E,E'} \sum_{a,a'} V_{aa'} G_a(E) G_a(-E) \left[ G_a (E) + G_a (-E) \right] G_{a'} (E'),
\ee
where the matrix element of the interaction $V_{aa'} = - (1 - s) (\lambda / \nu) \int d\bm{r} \, |\phi_a(\bm{r})|^2 |\phi_{a'}(\bm{r})|^2$ includes both Fock (exchange) and Hartree terms (with spin degeneracy factor $s=2$ in the latter). Here $\phi_a(\bm{r})$ are the wavefunctions corresponding to the energies $\xi_a$.
The Matsubara Green function in this representation is $G_{a}(E) = 1/ (\xi_a - iE)$. After some algebra, Eq.\ \eqref{P_elastic} can be rewritten \cite{MaekawaFukuyama_Localization2DSC_1982, Maekawa_UpperCriticalField2DSC_1983} in the form
\be
	\delta P^\text{elastic} =  T^2 \sum_E \frac{1}{2 i E}  \sum_{E'} \sum_{a a'} V_{aa'} \left[ G_a^2(E) - G_a^2(-E) \right] G_{a'} (E') ,
\ee
where one recognises corrections to the Green functions at coincident points ($\delta G_E$ and $\delta G_{-E}$) summed over Matsubara energies with a factor $T / (2 i E)$. Adding renormalised vertices $\upsilon(E)$ [Eq.~\eqref{upsilon}] and substituting to Eq.~\eqref{dTc_through_Pi}, one arrives at the expression
\be
	\frac{ \delta T_{c}^{\text{elastic}} }{T_c} = \left( \frac{\lambda_{\text{ph}}}{\lambda_{\text{BCS}}} \right)^2 iT \sum_{E=E_n} \frac{u(E)^2}{E} \frac{\delta G_{E} - \delta G_{-E}} {\nu_0} .
\ee
Now using the analyticity property, which relates the Matsubara Green function with the real-time retarded Green function $G^{R}$ (at coinciding points in our case),
\be
	G_E = \frac{1}{\pi} \int d \varepsilon \, \frac{\Im G^{R} (\varepsilon)  }{\varepsilon - i E} = -\int d \varepsilon \, \frac{\nu (\varepsilon)  }{\varepsilon - i E} . 
\ee
One can finally express \cite{Belitz-correlation-gap} the result for the contribution of elastic diagrams to the $T_c$ shift via the correction to the tunneling density of states, arriving at Eq.~\eqref{elastic_via_DOS}.

\end{document}